\documentclass[preprintnumbers,article,amsmath,amssymb,floatfix,10pt,prd,onecolumn,superscriptaddress,nofootinbib]{revtex4-2}

\usepackage{titlesec}
\usepackage[toc]{appendix}
\usepackage{caption}
\titleformat*{\section}{\LARGE\bfseries}
\titleformat*{\subsection}{\Large\bfseries}
\titleformat*{\subsubsection}{\large\bfseries}
\titleformat*{\paragraph}{\large\bfseries}
\titleformat*{\subparagraph}{\large\bfseries}
\usepackage{bm}
\usepackage{amsfonts}
\usepackage{latexsym}
\usepackage[latin1]{inputenc}
\usepackage{graphicx}
\usepackage{amsmath}
\usepackage{palatino}
\usepackage{mathpazo}
\usepackage{textcomp}
\usepackage{float}
\usepackage{booktabs}
\usepackage{dcolumn}
\usepackage{ragged2e}
\usepackage{hyperref}
\hypersetup{colorlinks,citecolor=blue}
\hypersetup{colorlinks=true,linkcolor=blue,filecolor=magenta,urlcolor=blue}
\usepackage{amsthm}
\usepackage{xcolor}
\usepackage{orcidlink}
\usepackage{epsfig}
\usepackage{commath}
\usepackage{cancel, ulem}
\usepackage{subcaption}
\usepackage{caption}
\usepackage{multirow}
\newcommand{\m}{\mathring}
\newcommand{\be}{\begin{equation}}
\newcommand{\ee}{\end{equation}}
\newcommand{\bea}{\begin{eqnarray}}
\newcommand{\eea}{\end{eqnarray}}
\newcommand{\eeas}{\end{eqnarray*}}
\newcommand{\beas}{\begin{eqnarray*}}

\def\jnl@style{\it}
\def\aaref@jnl#1{{\jnl@style#1}}

\def\aaref@jnl#1{{\jnl@style#1}}

\def\aj{\aaref@jnl{AJ}}                   
\def\apj{\aaref@jnl{ApJ}}                 
\def\apjl{\aaref@jnl{ApJ}}                
\def\apjs{\aaref@jnl{ApJS}}               
\def\apss{\aaref@jnl{Ap\&SS}}             
\def\aap{\aaref@jnl{A\&A}}                
\def\aapr{\aaref@jnl{A\&A~Rev.}}          
\def\aaps{\aaref@jnl{A\&AS}}              
\def\mnras{\aaref@jnl{Mon.~Not.~Roy.~Astron.~Soc.}}             
\def\prd{\aaref@jnl{Phys.~Rev.~D}}        
\def\prc{\aaref@jnl{Phys.~Rev.~C}}  
\def\prl{\aaref@jnl{Phys.~Rev.~Lett.}}    
\def\qjras{\aaref@jnl{QJRAS}}             
\def\skytel{\aaref@jnl{S\&T}}             
\def\ssr{\aaref@jnl{Space~Sci.~Rev.}}     
\def\zap{\aaref@jnl{ZAp}}                 
\def\nat{\aaref@jnl{Nature}}              
\def\aplett{\aaref@jnl{Astrophys.~Lett.}} 
\def\apspr{\aaref@jnl{Astrophys.~Space~Phys.~Res.}} 
\def\physrep{\aaref@jnl{Phys.~Rep.}}      
\def\physscr{\aaref@jnl{Phys.~Scr}}       
\def\commat{\aaref@jnl{Comm.~Math.~Phys.}}              
\def\science{\aaref@jnl{Science}}               
\def\cqg{\aaref@jnl{Classical Quant.~Grav.}}            
\def\jpcs{\aaref@jnl{JPCS}}                                     
\def\ijmpd{\aaref@jnl{Int.~J.~Mod.~Phys.~D}}                    
\def\grg{\aaref@jnl{Gen.~Relat.~Gravit.}}               
\def\rpp{\aaref@jnl{Rep.~Prog.~Phys.}}          
\def\npa{\aaref@jnl{Nucl.~Phys.~A}}        
\def\lrr{\aaref@jnl{Living Rev.~Rel.}}                   
\def\jcap{\aaref@jnl{J.~Cosmology Astropart.~Phys.}}    
\def\rmp{\aaref@jnl{Rev.~Mod.~Phys.}}   
\def\epjc{\aaref@jnl{Eur.~Phys.~J.~C}} 
\def\plb{\aaref@jnl{~Phy.~Lett.~B}} 
\def\mpla{\aaref@jnl{Mod.~Phy.~Lett.~A}} 
\def\arxiv{\aaref@jnl{arxiv.org}}

\allowdisplaybreaks[1]

\begin{document}

\title{Density contrast in the scalar-tensor extension of non-metricity gravity}
\author{Ganesh Subramaniam\orcidlink{0000-0001-5721-661X}}
\email{ganesh03@1utar.my}
\affiliation{Department of Mathematical and Actuarial Sciences, Universiti Tunku Abdul Rahman, Jalan Sungai Long,
43000 Cheras, Malaysia}
\author{Avik De\orcidlink{0000-0001-6475-3085}}
\email{avikde@um.edu.my}
\affiliation{Institute of Mathematical Sciences, Faculty of Science, Universiti Malaya, 50603 Kuala Lumpur, Malaysia}

\author{Jackson Levi Said\orcidlink{0000-0002-7835-4365}}
\email{jackson.said@um.edu.mt}
\affiliation{Institute of Space Sciences and Astronomy, University of Malta, Msida, Malta}

\begin{abstract}
We present a novel derivation of scalar cosmological perturbations in the scalar-tensor extension of non-metricity gravity, where the non-metricity scalar $Q$ is non-minimally coupled to a dynamical scalar field. While previous investigations of symmetric teleparallel gravity focused primarily on background evolution or specialised gauge choices, a complete treatment of scalar perturbations in this non-minimally coupled framework has remained unexplored. In this work, we derive the full set of perturbed field equations, impose the quasi-static approximation, and obtain the effective Poisson equation together with the corresponding modified gravitational constant $G_{\rm eff}$. These ingredients allow us to construct the density contrast evolution equation and analyse the matter growth rate and growth index. Through numerical analysis, we showed that the scalar non-metricity theory is comparable to the well-known $\Lambda CDM$ model to some extent. The results provide a foundation for testing scalar non-metricity theories against large-scale structure observations and open new avenues for constraining non-minimally coupled non-metricity cosmologies.
\end{abstract}

\maketitle

\tableofcontents
\section{Introduction}\label{sec00}

Recent advances in science and technology have generated an unprecedented volume of high-precision observational data, revealing that the widely accepted Lambda Cold Dark Matter ($\Lambda$CDM) model provides an incomplete description of cosmic evolution. In addition to longstanding theoretical challenges such as the initial singularity, the requirement of an inflationary phase, the horizon problem, and the cosmological constant problem; recent observational discrepancies have become increasingly prominent. In particular, tensions in the measured values of the Hubble constant $H_0$ and the parameter $S_8$ highlight significant and intriguing anomalies that warrant further investigation \cite{valentino2021}. The Hubble tension $H_0$ problem is that the measurements of the present value of the accelerated expansion rate of the universe, reported by the early universe probes \cite{aghanim2020, ade2016} and the late universe probes \cite{riess2019, shajib2023}, exhibit discrepancies. The $S_8$ problem is the discrepancy between the predicted and observed matter clustering in the universe \cite{abdalla2022}. Furthermore, another recently identified problem is the varying dark energy or the cosmological constant, which was previously assumed to be constant \cite{notari2025}. 

In search of solving these problems, the researchers attempted to modify the geometric components of Einstein's field equation (EFE) in the theory of general relativity (GR). Concerning this, theories like $f(R)$, which is an extension of GR by replacing the Ricci scalar $R$ by a viable function of it in the Einstein-Hilbert action \cite{buchdahl1970}, Scalar tensor theory involving a non-minimal coupling of the Ricci scalar $R$ and a scalar field $\phi$ or its functional forms \cite{gao2014, gleyzes2015, gleyzes2015b, langlois2016} and teleparallel gravity, which ascribes gravity to torsion or non-metricity with zero curvature \cite{jimenez2019} etc., have been introduced. In the absence of both curvature and torsion, gravity is attributed to non-metricity, which leads to symmetric teleparallel gravity. Furthermore, a scalar field can enter the gravitational action in symmetric teleparallel gravity, establishing a scalar tensor extension of non-metricity theories \cite{jarv2018, murtaza2025b}.


As the most general scalar-tensor theory, Horndeski gravity has been extended to the framework of teleparallel gravity in \cite{bahamonde2019}. The authors demonstrated that the inclusion of the torsion tensor enriches the phenomenology of the theory, leading to a broader range of dynamical behaviours. In \cite{kadam2022}, the authors studied teleparallel gravity in the realm of scalar-tensor theories and discuss the stability of the critical points and their evolutionary behaviour. They found four extra critical points that could describe the dark energy era; two of them are stable de-Sitter solutions. In the newly proposed scalar non-metricity gravity, the extra scalar degree of freedom arising from the second connection class was investigated through phase space analysis \cite{murtaza2025}. A hierarchy of critical points including matter-dominated, stiff-fluid, and de Sitter solutions, along with trajectories leading to Big Crunch/Rip singularities and transient, unstable matter phases were found. The authors in \cite{murtaza2025a} studied scalar non-metricity gravity with steep potential and showed that under specific parametric conditions, the nonminimally coupled scalar-tensor non-metricity gravity can explain the early and late time cosmic expansions. Their results show that the time-dependent scalar and steep potential in non-metricity gravity can drive cosmic evolution and offer testable predictions for future surveys. 

Apart from background studies, cosmological perturbation theory plays a crucial role in bridging the early fluctuations in the energy density of the universe to the formation of structures, such as clusters of galaxies \cite{kodama1984, mukhanov1992, malik2009}. From the observed CMB at redshift $z\approx 1000$, the temperature fluctuation is about $\delta \text{T/T}\approx 10^{-5}$ which is of the same order as the fluctuation of the baryonic density $\delta\rho/\bar{\rho}$ \cite{plank2018}. From this time ($z\approx 1000$) onward, the growth of density perturbation up to the linear regime is described by the well-known density contrast evolution equation \cite{peebles1980, gleyzes2016} 
\[\ddot{\delta}(t)+2H(t)\dot{\delta}(t)=4\pi G\bar\rho\delta(t),\]\label{eeq}
where $H(t)$ is the Hubble parameter, $\bar\rho$ is the background energy density, and $G$ is the Newtonian constant. Although an initial perturbation of order $10^{-5}$ is sufficient to describe the formation of structures as observed today, the growth of the factor of $10^3$ since the time of recombination leads to the perturbation of order $10^{-2}$, which is smaller than the nonlinear structures observed at the present \cite{ostriker1993}. This is one of the key reasons that motivate researchers to explore perturbation theory in modified gravity theories. The Newtonian constant $G$ of equation (\ref{eeq}) is replaced by the effective gravitational constant $G_{eff}$ to indicate the modifications in the respective modified gravity theories, e.g., in $f(R)$ gravity \cite{tsujikawa2007, tsujikawa2008,dombriz2008, mirzatuny2014}. 
Cosmological perturbations have been studied in the realm of teleparallel and symmetric teleparallel gravity as well. The authors studied $f(T)$ gravity and scalar-torsion gravity and found that there are no extra degrees of freedom in the flat FLRW background in $f(T)$ models \cite{golovnev2018}. The evolution of the matter spectrum for scalar-torsion gravity was also provided. For a broader review of teleparallel gravity, in both theoretical and observational aspects, please refer to \cite{bahamonde2023} and the references therein. As well-known, the growth of structures plays an important role in validating modified gravity theories. In $f(T)$ gravity, it was found that the perturbation grows slower compared to general relativity when $f_T>0$ due to effectively weakened gravity \cite{zheng2011, capozziello2024}. 
In the $f(T, B)$ gravity involving the boundary term $B$, the $G_{eff}$ term was obtained  by adopting a Newtonian gauge using quasi-static and subhorizon approximations \cite{bahamonde2021}. In the symmetric teleparallel gravity and its extensions, cosmological perturbations were studied extensively \cite{jimenez2020,atayde2021, albuquerque2022, heisenberg2024}. Unfortunately, in most of these works, the authors considered a specialised gauge choice, i.e., Newtonian gauge, except the very recent work \cite{ganesh2024} in which a gauge-free scalar perturbation setup was derived and the corresponding density contrast equation and growth structure were analysed. Having $G_{eff}$ in the respective theories of modified gravity is important for studying the formation of structures from matter to the present dark energy-dominated era of the universe and for classifying the viable models of the respective modified gravity theories that confront observations. In addition to that, with the wealth of observational datasets, we could constrain the viable modified theory base. 


The paper is organized as follows: first, in section \ref{sec4}, we discuss the fundamentals of scalar non-metricity gravity in background spacetime; then, in section \ref{sec6}, we discuss the scalar cosmological perturbation for scalar non-metricity gravity; after that, in section \ref{dgi}, we analyze the density growth and growth index; and finally, in section \ref{conc}, we summarize our results.
\section{Background spacetime in scalar non-metricity gravity}\label{sec4}
In the symmetric teleparallel formulation of gravity, spacetime curvature and torsion are identically zero, and gravitational interactions are entirely encoded in the non-metricity of a flat, symmetric affine connection $\Gamma^\lambda{}_{\mu\nu}$. This connection satisfies the following conditions:
\begin{align}
\mathbb T^\alpha{}_{\mu\nu}:=&
2\Gamma^{\alpha}{}_{[\nu\mu]}=0\,,
\label{eqn:torsion-free}\\
R^\lambda{}_{\mu\alpha\nu}:=&
2\partial_{[\alpha}\Gamma^\lambda{}_{|\mu|\nu]}
+2\Gamma^\lambda{}_{\sigma[\alpha}\Gamma^\sigma{}_{|\mu|\nu]}\,.
\label{eqn:curvature-free}
\end{align}

The vanishing curvature implies path-independent parallel transport, hence the term ``teleparallel", while the absence of torsion ensures symmetry in the lower indices of $\Gamma^\lambda{}_{\mu\nu}$, justifying the descriptor ``symmetric". Under these conditions, the affine connection can be splitted into the Levi-Civita connection $\mathring{\Gamma}^\lambda{}_{\mu\nu}$ and a disformation tensor $L^\lambda{}_{\mu\nu}$:
\begin{equation} \label{connc}
\Gamma^\lambda{}_{\mu\nu} := \mathring{\Gamma}^\lambda{}_{\mu\nu}+L^\lambda{}_{\mu\nu}\,.
\end{equation}
The disformation tensor quantifies the deviation from metric compatibility and is algebraically related to the non-metricity tensor
\begin{equation} \label{Q tensor}
Q_{\lambda\mu\nu} := \nabla_\lambda g_{\mu\nu} = \partial_\lambda g_{\mu\nu} - \Gamma^{\beta}{}_{\lambda\mu} g_{\beta\nu} - \Gamma^{\beta}{}_{\lambda\nu} g_{\beta\mu} \neq 0\,,
\end{equation}
via
\begin{equation} \label{L}
L^\lambda{}_{\mu\nu} = \tfrac{1}{2}\left(Q^\lambda{}_{\mu\nu} - Q_\mu{}^\lambda{}_\nu - Q_\nu{}^\lambda{}_\mu\right)\,.
\end{equation}
From $Q_{\lambda\mu\nu}$, two independent non-metricity vectors are constructed using the inverse metric:
\begin{equation}
Q_\mu := g^{\nu\lambda} Q_{\mu\nu\lambda} = Q_\mu{}^\nu{}_\nu\,, \qquad 
\tilde{Q}_\mu := g^{\nu\lambda} Q_{\nu\mu\lambda} = Q_{\nu\mu}{}^\nu\,. \label{qvec}
\end{equation}
Analogously, the disformation vectors yields
\begin{align}
L_\mu := L_\mu{}^\nu{}_\nu\,, \qquad 
\tilde{L}_\mu := L_{\nu\mu}{}^\nu\,.
\end{align}

The gravitational dynamics are governed by a scalar-non-metricity coupling, for which the non-metricity conjugate (or superpotential) is defined as
\begin{equation} \label{P}
P^\lambda{}_{\mu\nu} = \frac{1}{4}\left(-2 L^\lambda{}_{\mu\nu} + Q^\lambda g_{\mu\nu} - \tilde{Q}^\lambda g_{\mu\nu} - \delta^\lambda{}_{(\mu} Q_{\nu)}\right)\,.
\end{equation}
Contracting this with the non-metricity tensor yields the scalar invariant
\begin{equation} \label{Q}
Q = Q_{\alpha\beta\gamma} P^{\alpha\beta\gamma}\,,
\end{equation}
which serves as the fundamental geometric Lagrangian density in non-metricity-based gravity.

Extending this framework with a minimally coupled scalar field $\phi$, we consider the action \cite{jarv2018}
\begin{align}\label{eqn:ST}
S=\frac{1}{2\kappa^2 }\int\sqrt{-g}\left[f(\phi)Q-h(\phi)\nabla^\alpha\phi\nabla_\alpha\phi-U(\phi)
+2\kappa^2\mathcal L_m \right] \,d^{4}x\,,
\end{align}
where $U(\phi)$ is the scalar potential, and $f(\phi)$ governs the non-minimal coupling to the non-metricity scalar $Q$. As in standard scalar-tensor theories, the action is invariant under reparametrizations of $\phi$, implying one of the arbitrary functions $f$ or $h$ can be fixed by a field redefinition.

Varying the action with respect to the metric $g^{\mu\nu}$ yields the metric field equations:
\begin{align}
\kappa^2 T_{\mu\nu}
=&f\m G_{\mu\nu} +
2f'P^\lambda{}_{\mu\nu}\nabla_\lambda \phi
-h\nabla_\mu\phi\nabla_\nu\phi
+\frac12hg_{\mu\nu}\nabla^\alpha\phi\nabla_\alpha\phi+\frac12Ug_{\mu\nu} \,,
\label{eqn:FE1}
\end{align}
where $\mathring{G}_{\mu\nu}$ is the Einstein tensor built from the Levi-Civita connection, and the stress-energy tensor is defined variationally as
\begin{align*}
T_{\mu\nu}=-\frac 2{\sqrt{-g}}\frac{\delta(\sqrt{-g}\mathcal L_M)}{\delta g^{\mu\nu}}\,.
\end{align*}
 In the present paper, we consider a perfect fluid type stress energy tensor given by
\begin{align}
T_{\mu\nu}=pg_{\mu\nu}+(p+\rho)u_\mu u_\nu\,,
\end{align}
where $\rho$, $p$ and $u^\mu$ denote 
the energy density, pressure, and four velocity of the fluid, respectively.
On the other hand, the scalar field equation, obtained by varying the action with respect to $\phi$, reads:
\begin{align}\label{eqn:FE2}
f'Q+h'\nabla^\alpha\phi\nabla_\alpha\phi+2h\m\nabla^\alpha\m\nabla_\alpha \phi-U'=0\,,
\end{align}
where primes denote derivatives with respect to $\phi$. 
Finally, variation with respect to the affine connection (assuming vanishing hypermomentum) gives the connection field equations:
\begin{align}\label{eqn:FE3}
(\nabla_\mu-\tilde L_\mu)(\nabla_\nu-\tilde L_\nu)
\left[fP^{\mu\nu }{}_\lambda\right]=0\,,
\end{align}
in the absence of the hypermomentum tensor.


Taking the covariant divergence of (\ref{eqn:FE1}) and using (\ref{eqn:FE2}) and (\ref{eqn:FE3}) yields the conservation law for matter:
\begin{align}
    \mathring\nabla_\alpha T^\alpha{}_{\mu\nu}=0\,.\label{diveq}
\end{align}


Assuming the cosmological principle, we adopt the spatially flat Friedmann-Lema\^{i}tre-Robertson-Walker (FLRW) metric:
\begin{align}\label{ds:RW}
ds^2=-dt^2+a^2\delta_{ij}dx^idx^j\,,
\end{align}
where $a(t)$ is the scale factor of the Universe. The background cosmological fluid is of the form of a perfect fluid $T_{\mu\nu}=(\bar\rho,a^2\bar p,a^2\bar p,a^2\bar p)$, where an overbar ($\bar{\quad}$) representing quantities in background spacetime.

From equation (\ref{eqn:FE1}), we obtained the modified Friedmann equations:
\begin{align}
    3H^2=&\kappa^2(\bar \rho+\bar \rho^{DE})\,,\label{fle1}\\
    -(2\dot{H}+3H^2)=&\kappa^2(\bar p+\bar p^{DE})\,,\label{fle2}
 \end{align}
 where $H=\dot{a}/a$, and the dark energy density and pressure are
\begin{align}
    \kappa^2\rho^{DE}=&3H^2(1-f)+\frac{h\dot{\phi}^2}{2}+\frac{U}{2}\,,\label{rode1}\\
    \kappa^2p^{DE}=&-(2\dot{H}+3H^2)(1-f)+2H\dot{f}+\frac{h\dot{\phi}^2}{2}-\frac{U}{2}\,.\label{pde1}
\end{align}
The scalar field equation (\ref{eqn:FE2}) simplified to
\begin{align}\label{eqn:FE2-1}
-6H^2f'-h'\dot \phi^2 -2h(\ddot\phi+3H\dot \phi)-U'=0\,,
\end{align}
while matter conservation follows from (\ref{diveq}):
\begin{align}
\dot{\bar \rho}+3 H(\bar\rho+\bar p)=0.
\end{align}

\begin{figure}[h!]
 \begin{minipage}{0.5\textwidth}
    \includegraphics[width=0.8\linewidth]{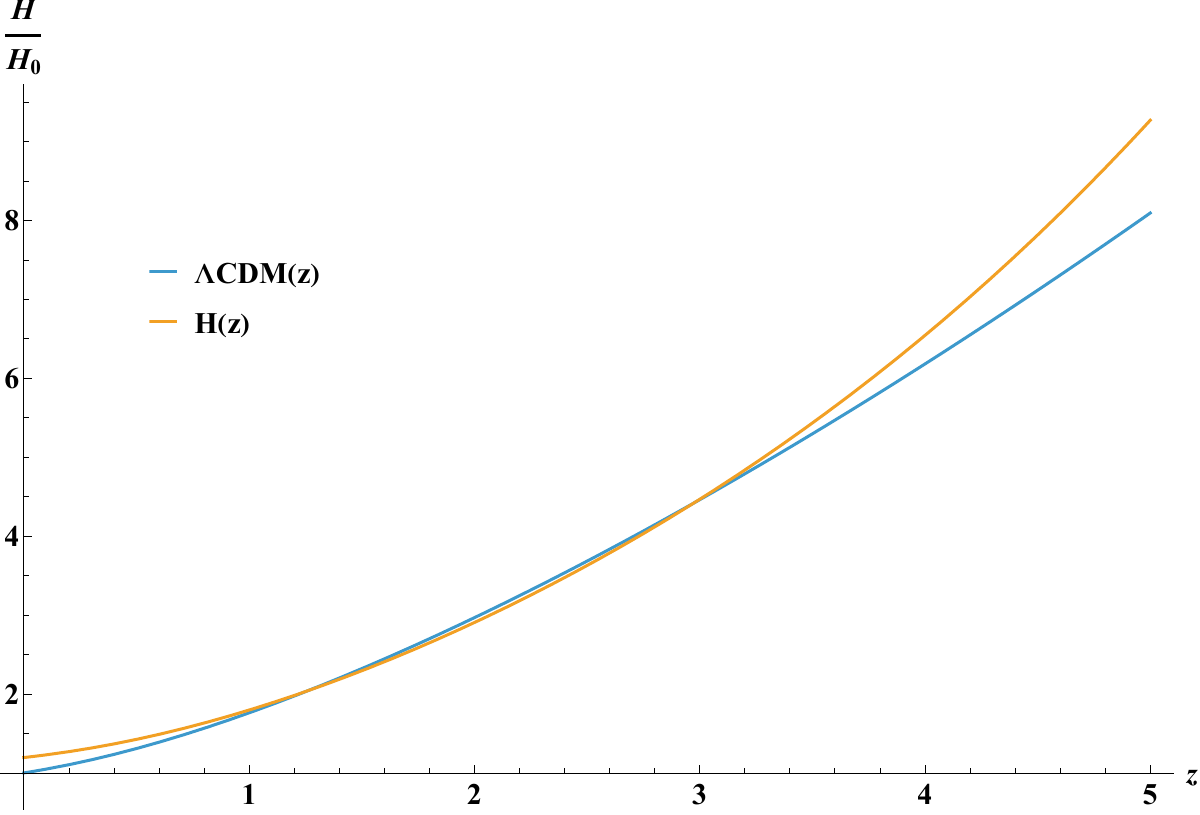}
    \caption{$H/H_0$ vs $z$, $n=0.21$, $f_0=10^{-8}$ and $U_0=0.7223$.}
    \label{hvsz}
 \end{minipage}%
 \begin{minipage}{0.5\textwidth}
    \includegraphics[width=0.8\linewidth]{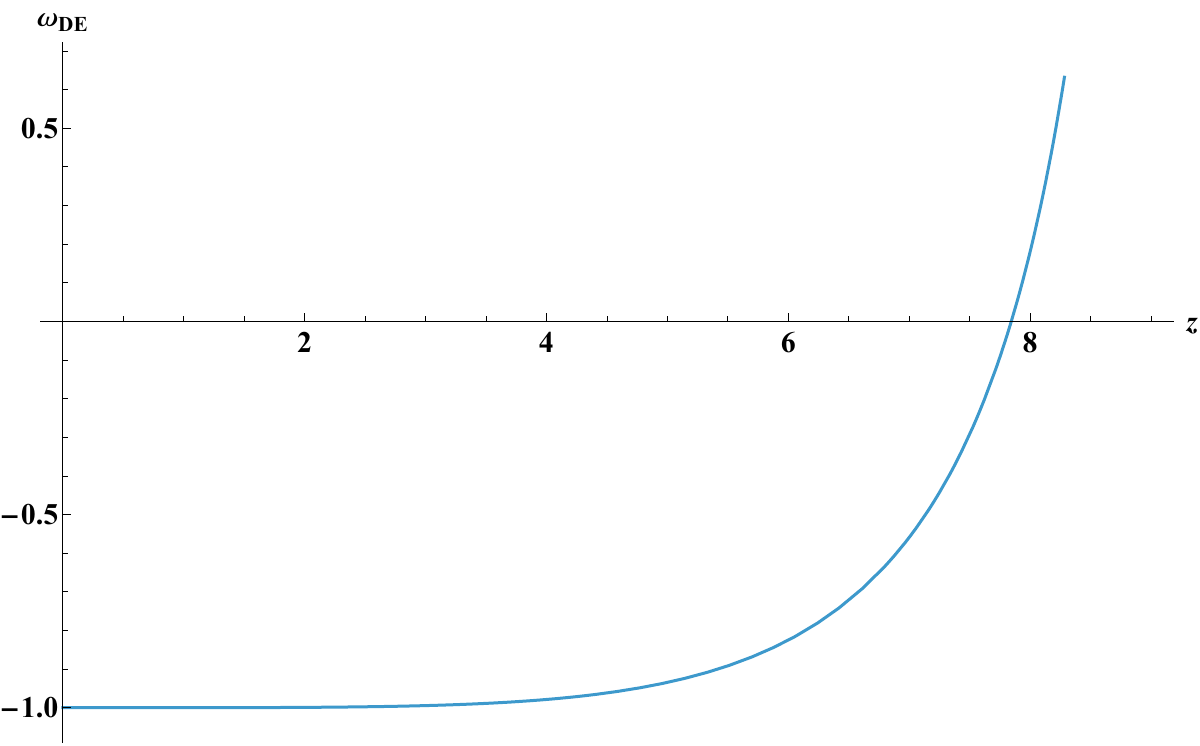}
    \caption{Evolution of Dark Energy equation of state parameter $\omega^{DE}$.}
    \label{eos}
 \end{minipage}
\end{figure}

As an illustrative model, we adopt
\[
f(\phi) = 1 + f_0 \phi^2\,, \qquad U(\phi) = U_0 \phi^n\,, \qquad  h(\phi) = h_0\,,
\]
and perform a numerical integration of the background equations. The resulting normalised Hubble parameter $H(z)/H_0$ matches observational trends and is compared with the $\Lambda$CDM prediction in figure \ref{hvsz} for the scalar non-metricity gravity compared with $\Lambda$CDM.

In the matter-dominated era ($\bar{p} = 0$), and assuming a dark energy equation of state $\bar{p}^{\mathrm{DE}} = \omega^{\mathrm{DE}} \bar{\rho}^{\mathrm{DE}}$ equation (\ref{fle2}) leads to
\begin{equation}
    -\frac{2\dot{H}}{3H^2}=\omega^{DE}(1-\Omega^{m})+1\,,\label{eeq1}
\end{equation}
where the matter density parameter is
\begin{equation}
    \Omega^m=\frac{\Omega^{m0}}{a^3}\left(\frac{H_0}{H}\right)^2\,,\label{om}
\end{equation}
with $\Omega^{m}_0$ and $H_0$ denoting present-day values.

The dark energy equation of state parameter $\omega^{DE}$ is derived from equation (\ref{rode1}) and (\ref{pde1}) as
\begin{equation}
    \omega^{DE}=\frac{\bar p^{DE}}{\bar \rho^{DE}}=-1-\frac{2 \left(-2 H \dot{\phi}f'-2 (f-1) \dot{H}-h_0 \dot{\phi}^2\right)}{-6 H^2 (f-1)+h_0 \dot{\phi}^2+U}\,.\label{wde}
\end{equation} 
The evolution of $\omega^{DE}$ is illustrated in figure \ref{eos}.

\section{Scalar cosmological perturbation}\label{sec6}
We start with the scalar metric perturbation (see Appendix \ref{A000} for details)
\begin{equation}
    ds^2=-(1+2\varphi)dt^2+2a \partial_iB dx^idt+a^2\left[(1-2\varsigma)\delta_{ij}+2\partial_{ij}E\right]dx^idx^j\,,\label{smp0}
\end{equation}
where $\varphi, \varsigma, B$ and $E$ are parameters depending on $(t,x^1,x^2,x^3)$.
Perturbation of field equation (\ref{eqn:FE1}) is
\begin{align}\label{pfe1}
    \kappa^2\delta T_{\mu\nu}=&f'\mathring{G}_{\mu\nu}\delta\phi+f\delta G_{\mu\nu}+2\left[f''P^\lambda{}_{\mu\nu}(\nabla_\lambda\phi)\delta\phi+f'\nabla_\lambda\phi\delta P^\lambda{}_{\mu\nu}+f' P^\lambda{}_{\mu\nu}\nabla_\lambda\delta\phi\right]-h'\nabla_\mu\phi\nabla_\nu\phi(\delta\phi)\notag\\&-h\left(\nabla_\mu\delta\phi\nabla_\nu\phi+\nabla_\mu\phi\nabla_\nu\delta\phi\right)+\frac{1}{2}\left[\left(h' g_{\mu\nu}\delta\phi+h\delta g_{\mu\nu}\right)\nabla^\alpha\phi\nabla_\alpha\phi\right.\notag\\&\left.+hg_{\mu\nu}\left(\nabla_\alpha\phi\nabla_\beta\phi\delta g^{\alpha\beta}+g^{\alpha\beta}(\nabla_\alpha\phi\nabla_\beta\delta\phi+\nabla_\alpha\delta\phi\nabla_\beta\phi)\right)+U' g_{\mu\nu}\delta\phi+U\delta g_{\mu\nu}\right]\,,
\end{align}
where the energy-momentum tensor perturbations for a perfect fluid is 
\begin{equation}
    \delta \mathcal{T}_{\mu\nu}=(\delta\rho+\delta p)\bar u_\mu\bar u_\nu+(\bar \rho+\bar p)(\bar u_\mu\delta u_\nu+\bar u_\nu\delta u_\nu)+\bar p\delta g_{\mu\nu}+\bar g_{\mu\nu}\delta p\,.\label{deltap}
\end{equation}
Here $\bar u_\mu$ and $\delta u_\mu$ are background fluid 4-velocity and its perturbation respectively. The 4-velocity perturbation is $\delta u_\mu=-\partial_\mu u$ where the vector contribution is ignored.
Hence, for the scalar metric perturbation, the components of the energy-momentum tensor perturbation (\ref{deltap}) are
\begin{align}\label{pemt1}
    \delta \mathcal{T}_{00}=&\delta\rho+2\bar{\rho}\varphi\,, 
    \quad\quad\quad\quad\quad\quad\quad\quad\quad\quad
    \delta \mathcal{T}_{0i}=\delta \mathcal{T}_{i0}=(\bar{\rho}+\bar{p})\partial_iu+a\bar p \partial_iB\,,
    \notag\\
    \sum_i\delta \mathcal{T}_{ii}=&3\left(2a^2\bar{p}\left(-\varsigma+\frac{1}{3}\nabla^2E\right)+a^2\delta p\right)\,,\quad\quad\delta \mathcal{T}_{ij}=2a^2\bar p\partial_{ij}E\,,
\end{align}
where an overbar $(\bar\quad)$ denotes quantities related to the background spacetime. Therefore, from the perturbation field equation (\ref{pfe1}), we have obtained the following perturbation quantities for the scalar-tensor extension of non-metricity gravity by taking $h(\phi)=h_0$ as a constant:
\begin{align}
    \kappa^2\delta\rho=&h_0(\ddot{\phi}+3H\dot{\phi})\delta\phi-h_0\dot{\phi}\dot{\delta\phi}+h_0\varphi\dot{\phi}^2\notag\\&+f'\left(6H^2\delta\phi-\dot\phi\frac{\nabla^2B}{2a}\right)+f\left[-6H(\dot{\varsigma}+H\varphi)+2H\nabla^2\dot{E}+2\left(\frac{\nabla^2\varsigma}{a^2}-H\frac{\nabla^2B}{a}\right)\right]\,,\label{speq1}\\
    \kappa^2(\bar\rho+\bar p)u=&2f(H\varphi+\dot{\varsigma})+\frac{f'}{2}\left[-H\delta\phi+\dot{\phi}(\varphi+3\varsigma-\nabla^2E)\right]\,,\label{speq2}\\
    \kappa^2\delta p=&h_0\dot{\phi}^2\varphi-h_0(\ddot{\phi}+3H\dot{\phi})\delta\phi-h_0\dot{\phi}\dot{\delta\phi}-3H^2f'\delta\phi-2H\dot{\phi}f''\delta\phi-f\left[-2(3H^2+2\dot{H})\varphi-2H(\dot{\varphi}+3\dot{\varsigma}-\nabla^2\dot{E})\right.\notag\\&\left.-2\ddot{\varsigma}+\frac{2}{3}\nabla^2\ddot{E}-\frac{2}{3}\frac{\nabla^2}{a}\left(\dot{B}+2HB+\frac{\varphi}{a}-\frac{\varsigma}{a}\right)\right]-f'\left[(3H^2+2\dot{H})\delta\phi-4H\dot{\phi}\varphi+2H\dot{\delta\phi}\right.\notag\\&\left.-2\dot{\phi}\left(\dot{\varsigma}-\frac{\nabla^2\dot{E}}{3}\right)-\dot{\phi}\frac{\nabla^2B}{6a}\right]\,,\label{speq3}\\
    -a^2\kappa^2\Pi=0=&f(\varphi-\varsigma+2aHB+a\dot{B}-3a^2H\dot{E}-a^2\ddot{E})+f'(aB-a^2\dot{E})\dot{\phi}\,.\label{speq4}
\end{align}
The perturbation of the second field equation (\ref{eqn:FE2}) is
\begin{align}\label{sfe}
    &f'\delta Q+Qf''\delta\phi+(h''g^{\alpha\beta}\delta\phi+h'\delta g^{\alpha\beta})\nabla_\beta\phi\nabla_\alpha\phi\\&+h'g^{\alpha\beta}\left(\nabla_\alpha\nabla_\beta\delta\phi+\nabla_\alpha\delta\phi\nabla_\beta\phi\right)\notag+2\left[(h'g^{\alpha\beta}\delta\phi+h\delta g^{\alpha\beta})\mathring{\nabla}_{\beta}\mathring{\nabla}_\alpha\phi+h g^{\alpha\beta}\delta(\mathring{\nabla}_\beta\mathring{\nabla}_\alpha\phi)\right]-U''\delta\phi=0\,,
\end{align}
where $\delta(\mathring{\nabla}_\beta\mathring{\nabla}_\alpha\phi)=\partial_{\alpha\beta}\delta\phi-(\delta\mathring{\Gamma}^\gamma{}_{\alpha\beta})\partial_\gamma\phi-\mathring{\Gamma}^\gamma{}_{\alpha\beta}\partial_\gamma\delta\phi$. For the scalar metric perturbation (\ref{smp0}), equation (\ref{sfe}) with $h(\phi)=h_0$ can be further simplified using $U''=-6H^2f''$ from equation (\ref{eqn:FE2-1}) as 
\begin{align}
    f'\delta Q+2h_0\left[2\varphi\ddot{\phi}+(3\dot{\varsigma}-\nabla^2\dot{E}+\dot{\varphi}+6H\varphi)\dot{\phi}-3H\dot{\delta\phi}-\ddot{\delta\phi}+\dot{\phi}\frac{\nabla^2B}{a}+\frac{\nabla^2\delta\phi}{a^2}\right]=0\,,\label{sfe01}
\end{align}
where $\delta Q=-2H\left[-6(\dot{\varsigma}+\varphi H)+2\nabla^2\dot{E}-\frac{\nabla^2B}{2a}\right]$. Furthermore, the perturbation of the third field equation (\ref{diveq}) is
\begin{align}
    \kappa^2\delta(\mathring{\nabla}_\mu T^\mu{}_\nu)=&2\left[\delta(\mathring{\nabla}_\mu P^{\lambda\mu}{}_\nu)+(\delta P^{\lambda\mu}{}_\nu)\mathring{\nabla}_\mu\right]\nabla_\lambda f+2\left(\mathring{\nabla}_\mu P^{\lambda\mu}{}_\nu+P^{\lambda\mu}{}_\nu \delta[\mathring{\nabla}_\mu\right)\nabla_\lambda f]\notag\\&+\left[\delta\mathring{G}^\lambda{}_\nu+\frac{\delta Q}{2}\delta^\lambda{}_\nu\right]\nabla_\lambda f+\left(\mathring{G}^\lambda{}_\nu+\frac{Q}{2}\delta^\lambda{}_\nu\right)\delta\left[\nabla_\lambda f\right]\,.\label{coneq}
\end{align}
From the fundamental law of conservation, we let $\delta(\mathring{\nabla}_\mu T^\mu{}_\nu)=0$, thence from equation (\ref{coneq}) the perturbation of energy and momentum conservation equation can be expressed as
\begin{align}
   &f'\left[\dot{\phi}\left(H\frac{\nabla^2B}{a}+\frac{\nabla^2\dot{B}}{2a}+\frac{\nabla^2\varphi}{a^2}-\frac{\nabla^2\varsigma}{2a^2}-\frac{\nabla^2\nabla^2E}{2a^2}\right)+\ddot{\phi}\frac{\nabla^2B}{2a}-H\frac{\nabla^2\delta\phi}{2a^2}\right]+ \dot{\phi}^2f''\frac{\nabla^2B}{2a}=0\,,\label{speceq1}\\
    &f'\left[\dot{\phi}\left(-2\dot{H}\frac{\nabla B}{a}+2H\frac{\nabla\varphi}{a^2}+2\frac{\nabla\dot{\varsigma}}{a^2}-\frac{\nabla^3B}{2a}-\frac{9}{2}H\nabla\varphi-\nabla\dot{\varphi}+\frac{\nabla\dot{\varsigma}}{2}+\frac{\nabla^3\dot{E}}{2}-\frac{9}{2}H\nabla\varsigma+\frac{3}{2}H\nabla^3E\right)\right.\notag\\&\left.-\frac{\ddot{\phi}}{2}\nabla(\varphi+3\varsigma-\nabla^2E)-\frac{3}{2}(3H^2+\dot{H})\nabla\delta\phi-\frac{3}{2}H\nabla\dot{\delta\phi}\right]+\frac{\dot{\phi}}{2}f''\left(-3H\nabla\delta\phi-\dot{\phi}\nabla(\varphi+3\varsigma-\nabla^2E)\right)=0\,.\label{speceq2}
\end{align}
According to the law of conservation, the perturbations of the conservation equation $\delta(\mathring{\nabla}^\mu T_{\mu\nu})=0$ for a perfect fluid gives us a general equation
\begin{align}\label{add1}
    \ddot{\delta}=&\left[3(2\omega-c_s{}^2)-2\right]H\dot{\delta}+3(\omega-c_s{}^2)\left[\dot{H}+H^2(2-3\omega)+\frac{c_s{}^2}{3(\omega-c_s{}^2)}\frac{\nabla^2}{a^2}\right]\delta\notag\\&+(1+\omega)\left[3(2-3\omega)H\dot{\varsigma}+3\ddot{\varsigma}+\frac{\nabla^2\varphi}{a^2}+(-2+3\omega)H\nabla^2\dot{E}-\nabla^2\ddot{E}\right.\notag\\&\left.+(1-3\omega)H\frac{\nabla^2B}{a}+\frac{\nabla^2\dot{B}}{a}\right]\,,
\end{align}
by combining both energy and momentum conservation equation.
For a matter-dominated universe, radiation density is ignored compared to matter density, so that we can set $c_s{}^2\equiv\frac{\delta P}{\delta \rho}\approxeq\frac{\bar p}{\bar \rho}=\omega=0$ and the equation (\ref{add1}) above reduces to 
\begin{equation}\label{denp11}
\ddot{\delta}^m=-2H\dot{\delta}^m+6H\dot{\varsigma}+3\ddot{\varsigma}+\frac{\nabla^2}{a^2}\left[\varphi-2a^2H\dot{E}-a^2\ddot{E}+a(HB+\dot{B})\right]\,.
\end{equation}

\section{Density growth rate and growth index}\label{dgi}
In this section we employ quasi static (QS) approximation, where we neglect the time derivatives of scalar quantities in the scalar metric perturbation e.g. $\dot{\varphi}, \dot{\varsigma}, \dot{B}, \dot{E}$ to arrive at the final equations. We use the Fourier transform of the form
\begin{align}
    \varphi(\vec{x},t)=\int\frac{d^3\vec{k}}{\sqrt{2\pi}^3}\tilde{\varphi}_{\vec{k}}(t) e^{i\vec{k}\cdot \vec{x}}\,; \ \ \ \  \varsigma(\vec{x},t)=\int\frac{d^3\vec{k}}{\sqrt{2\pi}^3}\tilde{\varsigma}_{\vec{k}}(t) e^{i\vec{k}\cdot \vec{x}}\,,
\end{align}
and $\nabla^2\equiv -k^2$. Hence, using the QS approximation, equations (\ref{speq1}), (\ref{speq4}),(\ref{sfe01}), (\ref{speceq1}) and (\ref{speceq2}) can be respectively written as
\begin{align}
   &h_0(\ddot{\phi}+3H\dot{\phi})\tilde{\delta\phi}_{\vec{k}}+h_0\tilde{\varphi}_{\vec{k}}\dot{\phi}^2\notag\\&+f'\left(6H^2\tilde{\delta\phi}_{\vec{k}}+\dot\phi\frac{k^2\tilde{B}_{\vec{k}}}{2a}\right)+f\left[-6H^2\tilde{\varphi}_{\vec{k}}+2\left(-\frac{k^2\tilde{\varsigma}_{\vec{k}}}{a^2}+H\frac{k^2\tilde{B}_{\vec{k}}}{a}\right)\right]=\kappa^2\bar\rho\tilde{\delta}_{\vec{k}}\,,\label{QS1}\\
    &f(\tilde{\varphi}_{\vec{k}}-\tilde{\varsigma}_{\vec{k}}+2aH\tilde{B}_{\vec{k}})+a\tilde{B}_{\vec{k}}\dot{\phi}f'=0\,,\label{QS2}\\
    &-2Hf'\left(-6H\tilde{\varphi}_{\vec{k}}+\frac{k^2\tilde{B}_{\vec{k}}}{2a}\right)+2h_0\left(2\ddot{\phi}\tilde{\varphi}_{\vec{k}}+6H\dot{\phi}\tilde{\varphi}_{\vec{k}}-\dot{\phi}\frac{k^2\tilde{B}_{\vec{k}}}{a}-\frac{k^2\tilde{\delta\phi}_{\vec{k}}}{a^2}\right)=0\,,\label{sfe011}\\
    &f'\left[\dot{\phi}\left(\frac{H\tilde{B}_{\vec{k}}}{a}-\frac{\tilde{\varsigma}_{\vec{k}}-2\tilde{\varphi}_{\vec{k}}}{2a^2}+\frac{k^2\tilde{E}_{\vec{k}}}{2a^2}\right)+\ddot{\phi}\frac{\tilde{B}_{\vec{k}}}{2a}-H\frac{\tilde{\delta\phi}_{\vec{k}}}{2a^2}\right]+\dot{\phi}^2f''\frac{\tilde{B}_{\vec{k}}}{2a}=0\,,\label{QS3}\\
    &f'\left[\dot{\phi}\left(-2\dot{H}\frac{\tilde{B}_{\vec{k}}}{a}+\frac{k^2\tilde{B}_{\vec{k}}}{2a}+2H\frac{\tilde{\varphi}_{\vec{k}}}{a^2}-\frac{9}{2}H(\tilde{\varphi}_{\vec{k}}+\tilde{\varsigma}_{\vec{k}})\right)-\frac{\ddot{\phi}}{2}(\tilde{\varphi}_{\vec{k}}+3\tilde{\varsigma}_{\vec{k}})-\frac{3}{2}(3H^2+\dot{H})\tilde{\delta\phi}_{\vec{k}}\right]\notag\\&+\frac{\dot{\phi}}{2}f''\left[-3H\tilde{\delta\phi}_{\vec{k}}-\dot{\phi}(\tilde{\varphi}_{\vec{k}}+3\tilde{\varsigma}_{\vec{k}})\right]+\left[f'\left(-\frac{3}{2}H\dot{\phi}-\frac{\ddot{\phi}}{2}\right)-\frac{\dot{\phi}^2}{2}f''\right]k^2\tilde{E}_{\vec{k}}=0\,,\label{QS4}
\end{align}
where we have considered the time derivative of the scalar field perturbation $\dot{\delta \phi}, \ddot{\delta\phi}, \text{etc.}\sim 0$. This is to avoid the oscillation of the scalar field perturbation $\delta\phi$ when the wavenumber $k\rightarrow \infty$ in $k/a$ \cite{boisseau2000, felice2010}.
Using equations (\ref{bk}), (\ref{vasig}) and (\ref{fik}) from appendix \ref{A00} in equation (\ref{QS1}), in matter era we obtain the Poisson equation as
\begin{align}
    -\frac{k^2\tilde{\varphi}_{\vec{k}}}{a^2}=\left(2\frac{k^4G_1}{G_2}\right)\frac{\kappa^2}{2}\bar{\rho}^m\tilde{\delta}^m{}_{\vec{k}}\,,\label{poiseq}
\end{align}
where $G_1$ and $G_2$ are given in appendix \ref{A00}.

\begin{figure}[h!]
 \begin{minipage}{0.5\textwidth}
    \includegraphics[width=0.8\linewidth]{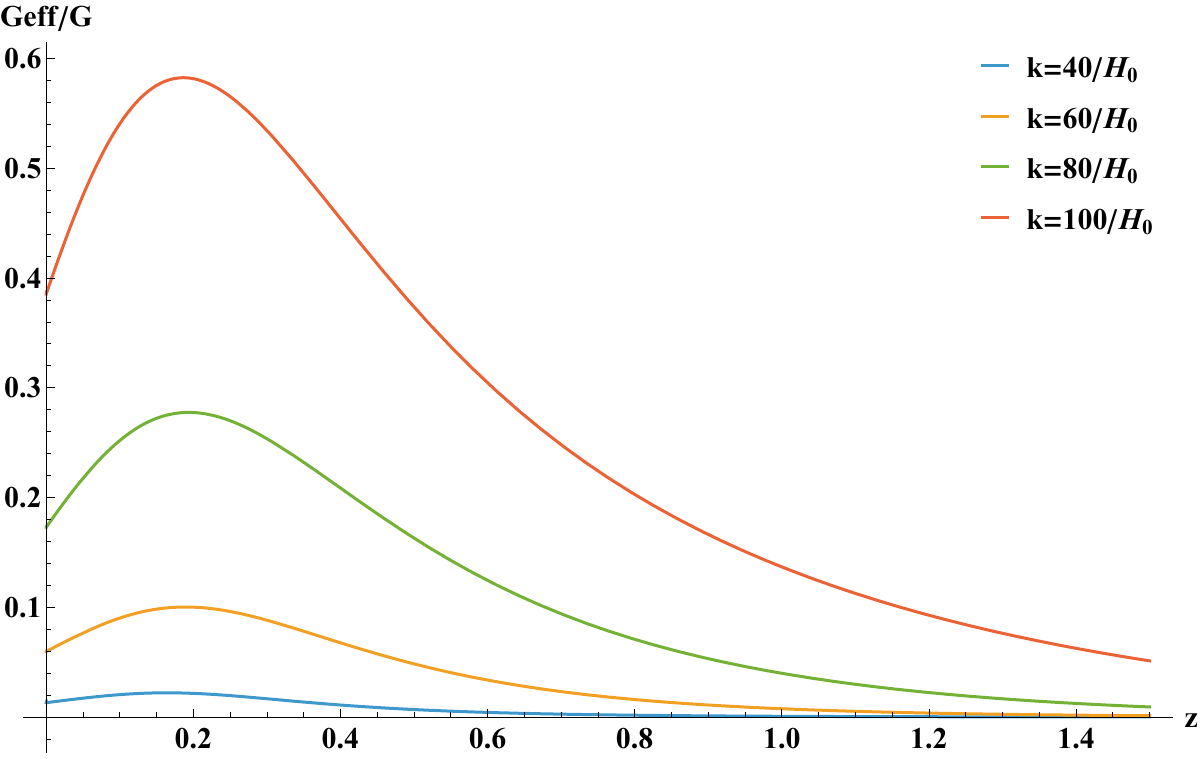}
    \caption{The ratio of effective gravitational constant $G_{ff}$ to that of Newtonian constant $G$, $G_{eff}/G$ vs redshift $z$.}
    \label{geff}
 \end{minipage}%
 \begin{minipage}{0.5\textwidth}
    \includegraphics[width=0.8\linewidth]{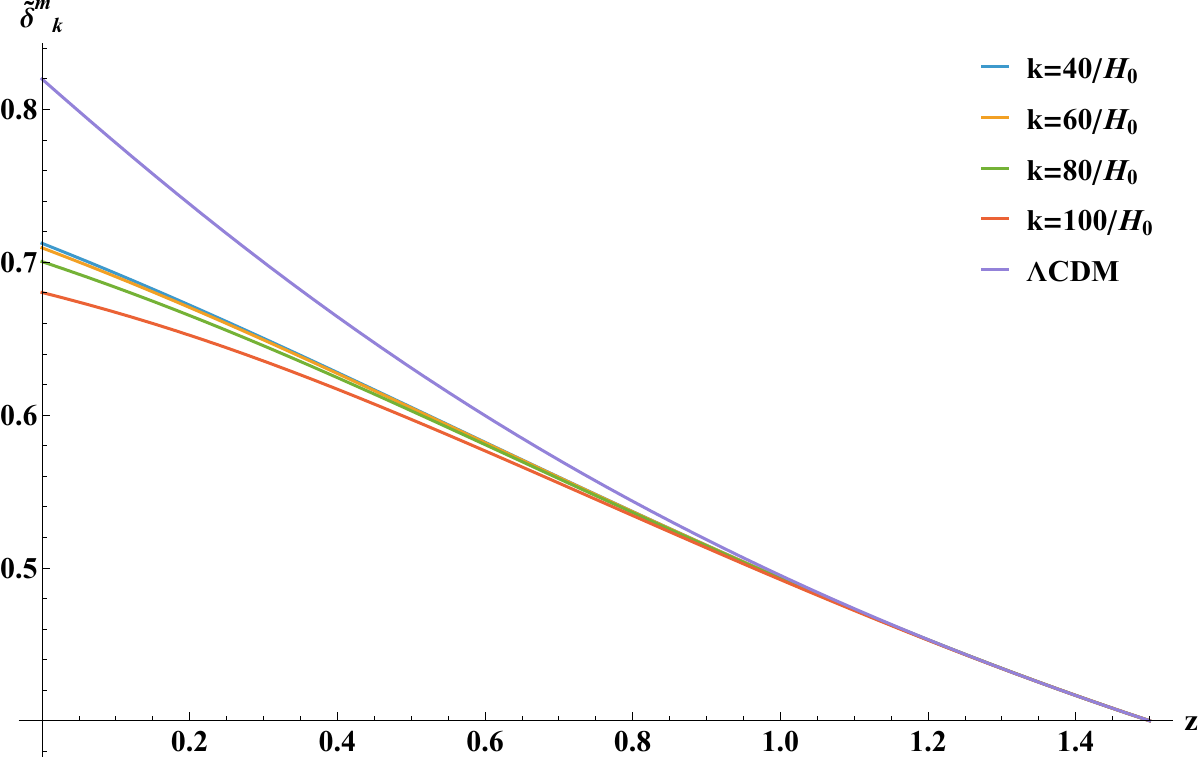}
    \caption{Matter density constrast $\tilde{\delta}^m{}_k$ vs redshift $z$ for several wavenumbers $k$. We took the initial condition for $\delta(z)=\frac{1}{1+z}$ and $\delta'(z)=-\frac{1}{(1+z)^2}$ at $z=1.5$.}
    \label{deltavsz}
 \end{minipage}
\end{figure}

The term $
    -\frac{k^2}{a^2}(\tilde{\varphi}_{\vec{k}}+aH\tilde{B}_{\vec{k}})$ in equation (\ref{denp11}) is similar to the poison equation (\ref{poiseq}) obtained earlier. Hence, using the QS approximation equation (\ref{denp11}) in the matter era can be expressed as
\begin{align}
    \ddot{\tilde{\delta}}^{m}{}_{\vec{k}}+2H\dot{\tilde{\delta}}^{m}{}_{\vec{k}}-4\pi G_{eff}\bar{\rho}^{m}\tilde{\delta}^{m}{}_{\vec{k}}=0\,,\label{deltaeq}
\end{align}
where $G_{eff}$, using equation (\ref{poiseq}) can be expressed\footnote{Note that only the QS approximation is considered to arrive at the $G_{eff}$.}
\begin{align}
    G_{eff}=\left(-2k^4\frac{G_1}{G_2}\right)G\,.\label{geff}
\end{align}
The behaviour of effective gravitational constant $G_{eff}$ against redshift $z$ is depicted in the figure \ref{geff} for different values of wavenumber $k$. For $\Lambda CDM$ model the $G_{eff}/G\approx 1$. Hence, from the figure \ref{geff} we observed that at wavenumber $k=100/H_0$, the ratio is $0.582$ at $z=0.2$ while for the other lesser wavenumbers the ratio deviating further lesser. Furthermore, the evolution of matter density contrast $\tilde{\delta}^m{}_k$ vs redshift $z$ for several wavenumbers are illustrated in figure \ref{deltavsz}. It is observed that the scalar non-metricity model considered is having comparable evolution to $\Lambda CDM$ at higher redshift $z$. While at the present time $z\approx 0$, the curve for wavenumbers $k=40/H_0$ and $60/H_0$ have closer value of about $0.71$ for matter density constrast $\tilde{\delta}^m{}_k$ comparable to the value of $0.82$ for the $\Lambda CDM$ model. Take note that this corresponds to a small value for the $G_{eff}/G$, marking the role of scalar non-metricity gravity for the cosmological application different than the $\Lambda CDM$ model.

\begin{figure}[h!]
 \begin{minipage}{0.5\textwidth}
    \includegraphics[width=0.8\linewidth]{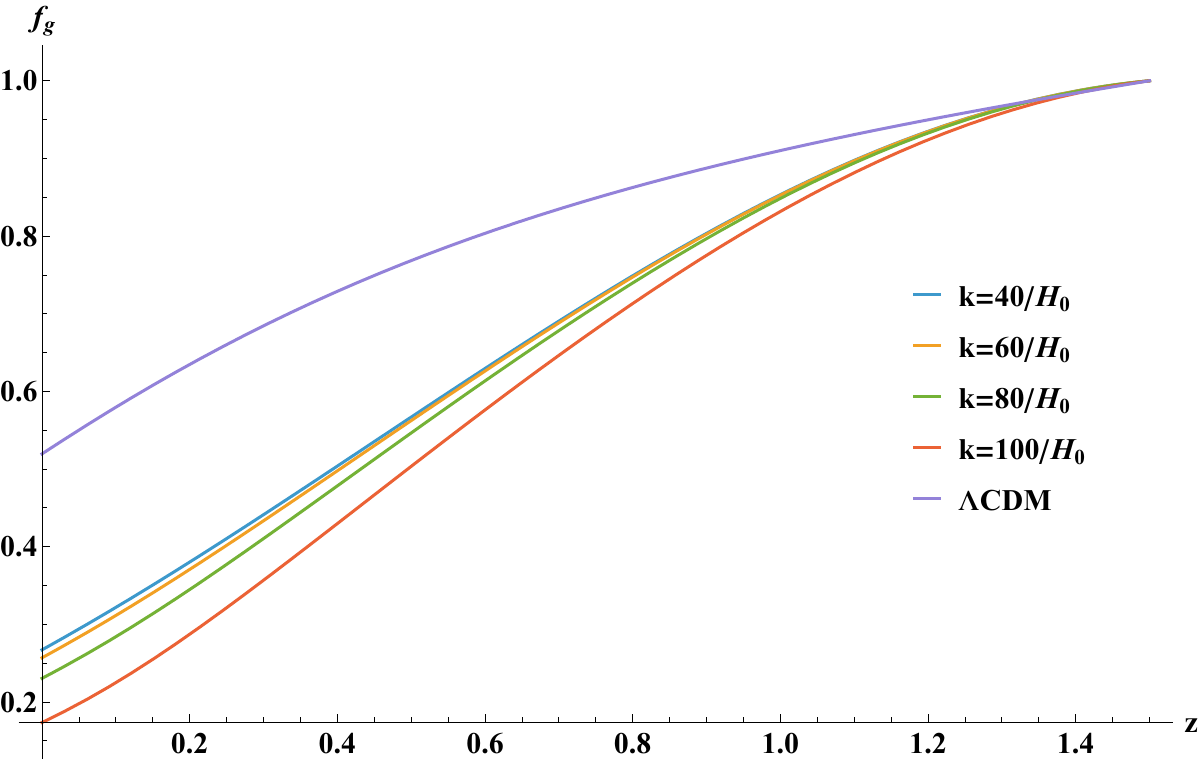}
    \caption{Evolution of matter growth rate $f_g$ vs redshift $z$ for various wavenumbers $k$.}
    \label{growth}
 \end{minipage}%
 \begin{minipage}{0.5\textwidth}
    \centering
    \includegraphics[width=0.8\linewidth]{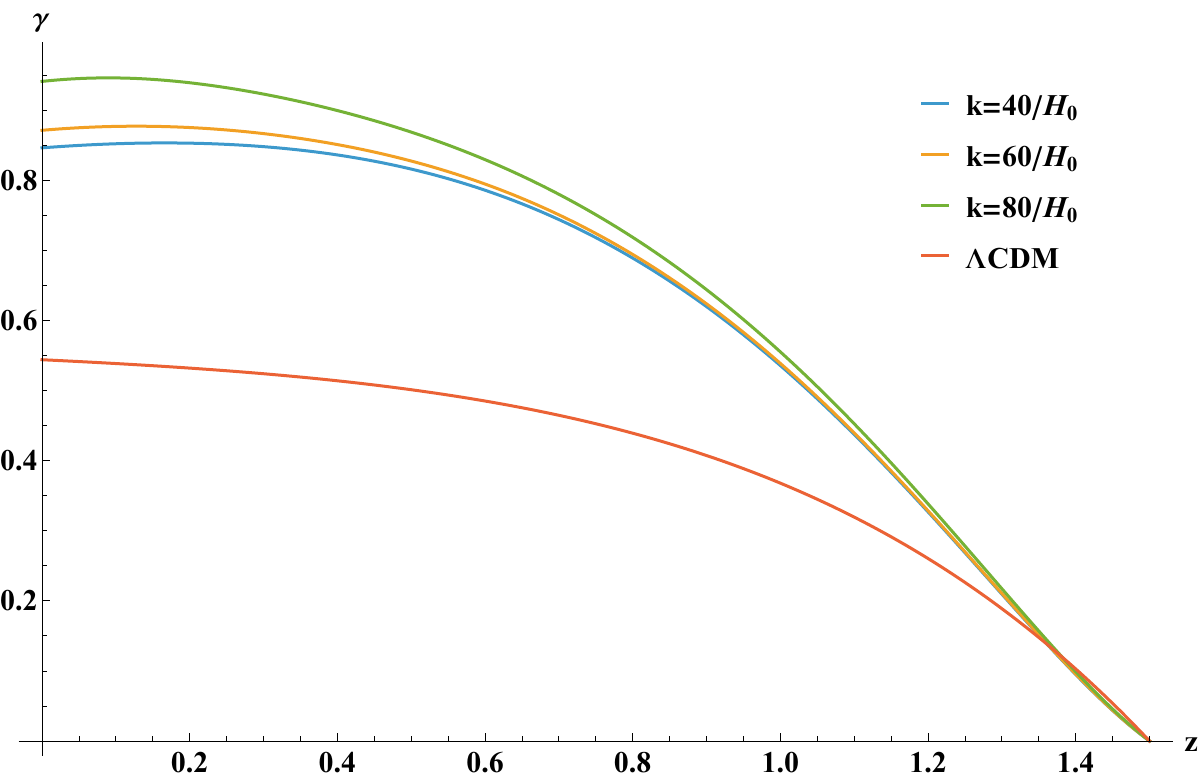}
    \caption{Evolution of growth index $\gamma$ vs redshift $z$.}
    \label{gamma}
 \end{minipage}
\end{figure}

We can utilise this to investigate the evolution of the matter growth rate $f_g$, a parameter that measures the growth of galaxies, clusters of galaxies, etc with respect to time. The matter growth rate $f_g$ is defined as \cite{peebles1993, linder2007, qing2014}
\begin{equation}
    f_g\equiv\frac{d\ln{\tilde{\delta}^m}{}_{\vec{k}}}{d\ln{a}}=-(1+z)\frac{\tilde{\delta}^{m^{'}}{}_{\vec{k}}(z)}{\tilde{\delta}^m{}_{\vec{k}}(z)}\,.\label{mg}
\end{equation}
By using the results obtained from equation (\ref{deltaeq}) which is illustrated in figure \ref{deltavsz}, we can plot the evolution of the matter growth rate $f_g$ against the redshift $z$ as depicted in figure \ref{growth}.
It can be observed from figure \ref{growth} that at the present ($z\approx 0$) the matter growth rate for scalar non-metricity gravity is $0.27$ at wavenumbers $k=40/H_0$ which is much lesser compared to the $\Lambda CDM$ model of $0.52$.


The most common parametrization of $f_g$ is in term of $(\Omega^m)^\gamma$ such that it is comparable to cosmological observations, which is written as \cite{peebles1993}
\begin{equation}
    f_g=(\Omega^m)^\gamma\,,\label{grop}
\end{equation}
where $\Omega^m$ and $\gamma$ are functions of redshift $z$. Here, $\gamma$ is the growth index of matter perturbation. The growth index $\gamma$ is a significant parameter to study the observed large-scale structure (LSS) of the universe to constrain different modified gravity models \cite{nesseris2015, basilakos2016, basilakos2020, kyllep2021, sharma2022, nguyen2023}. 


Equation (\ref{grop}) can be expressed as \cite{batista2014}
\begin{equation}
\gamma(z)=\frac{\log(f_g)}{\log(\Omega^m)}\,.
\end{equation}
Using equations (\ref{om}) and (\ref{mg}), we illustrate the evolution of the growth index $\gamma$ vs redshift $z$ in figure \ref{gamma}. It can be seen from the figure \ref{gamma}, the growth index for the curve with wavenumbers $k=40/H_0$ is $0.85$, higher than the $\Lambda CDM$ at $z=0$.
\section{Conclusion}\label{conc}

In this work, we developed the scalar cosmological perturbation theory for the scalar-tensor extension of non-metricity gravity with a non-minimally coupled scalar field. This represents the first systematic and fully consistent derivation of scalar perturbations within this formulation of symmetric teleparallel gravity. While previous studies in $f(Q)$ or non-minimally coupled teleparallel models have addressed perturbations in specific gauge choices or under restrictive assumptions, the non-minimally coupled non-metricity-scalar system analysed here required a careful treatment of the perturbed field equations, the perturbed non-metricity scalar, and the interplay between the scalar-field sector and geometric contributions. The derivation offered here fills an important gap in the literature and provides a complete foundation for studying structure formation in this class of theories.

We first analyzed the scalar non-metricity gravity theory by considering specific coupling and potential functions, $f=1+f_0 \phi(z)^2$ and $U=U_0\phi(z)^n$ respectively, and by assuming $f_0=10^{-8}, U_0=0.7223$ and $n=0.21$, we obtained the normalized Hubble parameter $H/H_0$ and the effective dark energy equation of state parameter $\omega^{DE}$, as depicted in figures \ref{hvsz} and \ref{eos}. This convinces us that the scalar non-metricity gravity theory is comparable to the $\Lambda CDM$ model of the universe at the background level. Having convincing numerical analysis results at the background level, we move on to analyse the perturbation of scalar non-metricity gravity theory.

By adopting the quasi-static approximation, we obtained the modified Poisson equation and extracted the effective gravitational constant $G_{\rm eff}$, which encapsulates the deviations from general relativity induced by the non-metricity geometry and scalar-field coupling. This is illustrated in figure \ref{geff}. It can be observed from the plot that the higher wavenumber has the ratio $G_{eff}/G$ closer to $1$ compared to the lower values of wavenumbers. However, when comparing the matter density contrast $\tilde{\delta}^m{}_{\vec{k}}$ with $\Lambda CDM$ as depicted in figure \ref{deltavsz}, the curve with wavenumber $k=40/H_0$ shows a difference of $0.11$. Furthermore, when comparing the growth rate $f_g$ of scalar non-metricity symmetric teleparallel gravity theory with $\Lambda CDM$ as depicted in figure \ref{growth}, it shows a deviation of $0.27$ for wavenumeber $k=40/H_0$ at $z=0$ which is about half orders of magnitude differences. Furthermore, the growth index $\gamma$ for the wavenumber $k=40/H_0$ is $0.85$ at $z=0$ which is larger than the value of $0.54$ for $\Lambda CDM$ model of the universe. The values obtained here can further enhanced by constraining the parameters, e.g. $f_0,U_0$ and $n$ based on the observational data set.


The results presented here naturally point toward several promising avenues for future research. A direct extension of this work would involve performing a full numerical evaluation of the growth index $\gamma(z)$ across a range of scalar-field potentials, coupling functions, and background evolutions, followed by a systematic comparison with current and upcoming large-scale structure datasets. Another direction is to generalise the perturbation analysis beyond the quasi-static limit, allowing for a more accurate treatment on near-horizon scales or in early-time cosmology. Moreover, the scalar perturbation framework developed here can be applied to study CMB anisotropies, matter power spectra, and weak lensing observables within minimally coupled non-metricity gravity. This would provide a comprehensive test of the theory against precision cosmological measurements.

Finally, the perturbative machinery constructed in this work can be extended to vector and tensor sectors, enabling a unified treatment of all cosmological perturbation modes. Such an analysis would open the possibility of examining gravitational-wave propagation, modified friction terms, and potential polarisation signatures arising from non-metricity geometry. Altogether, this study lays the groundwork for a complete phenomenological programme aimed at evaluating the viability of non-minimally coupled scalar-tensor non-metricity theories as candidates for explaining cosmic acceleration, structure formation, and possible deviations from general relativity.

\begin{acknowledgments}
AD and JLS acknowledge the Universiti Malaya BKP-ECRG Grant (Project No. BKP119-2025-ECRG). This article is also based upon work from COST Action CA21136 Addressing observational tensions in cosmology with systematics and fundamental physics (CosmoVerse) supported by COST (European Cooperation in Science and Technology). 
\end{acknowledgments}

\section{Appendix}
\appendix \label{appendix1}
\renewcommand{\theequation}{\thesection.\arabic{equation}}
\section{Metric Perturbation}\label{A000}
We consider first the general metric perturbation of the form\footnote{Note that the components are $g_{00}=-(1+2A), \ g_{0i}=g_{i0}=-aB_i, \ g_{ij}=a^2\left[(1+2D)\delta_{ij}+2E_{ij}\right]$.}
\begin{equation}\label{e1}
    ds^2=-(1+2A)dt^2-2a B_i dx^idt+a^2\left[(1+2D)\delta_{ij}+2E_{ij}\right]dx^idx^j\,,
\end{equation}
where $A=\varphi$, $D=-\varsigma+\frac{1}{3}\nabla^2E$ while $\varphi$ and $\varsigma$ are scalar functions of $(t,x^1,x^2,x^3)$.
It is common to perform scalar-vector-tensor (SVT) decomposition for $B_i$ and $E_{ij}$ \cite{kurkisuonio2024} \footnote{We followed \cite{kurkisuonio2024} notations, so our results might differ by a negative sign from some published works.}. First,

\begin{align}
    B_i=&B^S{}_i+\vec{B}_i=-\partial_iB+\vec{B}_i\,, \ \ \quad \nabla\times(B^S{}_i)=0\,, \ \ \ \ \nabla\cdot \vec{B}_i=0\,,\label{b}
\end{align}
where $B$ is a scalar function of $(t, x^1, x^2, x^3)$. Similarly,
\begin{align}
    E_{ij}=&E^S{}_{ij}+\vec{E}_{ij}+\vec{E}^T{}_{ij}\,,\label{e}
\end{align}
where $E^S{}_{ij}$ is the scalar term, $\vec{E}_{ij}$ the vector term and $\vec{E}^T{}_{ij}$ the tensor term which can be expressed as
\begin{align}
     E^S{}_{ij}=&\left(\partial_{ij}-\frac{1}{3}\delta_{ij}\nabla^2\right)E\,,\label{Es}\\
     \vec{E}_{ij}=&-\frac{1}{2}(\partial_i E_j+\partial_j E_i)\,, \ \ \ \ \delta^{ij}\partial_jE_i=\nabla\cdot\vec{E}=0\,,\\
     \delta^{ij}\partial_k E^T{}_{ij}=&0\,,\quad \delta^{ij}E^T{}_{ij}=0\,,
\end{align}
where $E$ is a scalar function of $(t, x^1, x^2, x^3)$.
In the following, we consider scalar metric perturbation. Hence, the vector and tensor contributions from $B_i$ and $E_{ij}$ are ignored. Therefore, by using $A=\varphi$, $D=-\varsigma+\frac{1}{3}\nabla^2 E$, and the $B^S{}_i$ , $E^S{}_{ij}$ terms of equations (\ref{b}) and (\ref{e}), the metric (\ref{e1}) can be written as
\begin{equation}
    ds^2=-(1+2\varphi)dt^2+2a \partial_iB dx^idt+a^2\left[(1-2\varsigma)\delta_{ij}+2\partial_{ij}E\right]dx^idx^j\,.\label{smp}
\end{equation}
\section{$G_1$ and $G_2$ of equation (\ref{geff})}\label{A00}
From equation (\ref{QS4}), we have
\begin{align}
k^2\tilde{E}_{\vec{k}}=&-\left(a^2\left(f' \left(\ddot{\phi }+3 H \dot{\phi }\right)+\dot{\phi }^2 f''\right)\right)^{-1}\notag\\&\times\left(f' \left(a^2 \left(\ddot{\phi } \left(3 \tilde{\varsigma }_{\vec{k}}+\tilde{\varphi }_{\vec{k}}\right)+3 \left(3 H^2+\dot{H}\right) \tilde{\delta \phi }_{\vec{k}}\right)+\dot{\phi } \left(9 a^2 H \tilde{\varsigma }_{\vec{k}}+\left(9 a^2-4\right) H \tilde{\varphi }_{\vec{k}}-a \left(k^2-4 \dot{H}\right) \tilde{B}_{\vec{k}}\right)\right)\right.\notag\\&\left.+a^2 \dot{\phi } f'' \left(3 H \tilde{\delta \phi }_{\vec{k}}+\dot{\phi } \left(3 \tilde{\varsigma }_{\vec{k}}+\tilde{\varphi }_{\vec{k}}\right)\right)\right)\,,\label{ek}
\end{align}
substitute into equation (\ref{QS3}), we have
\begin{align}
\tilde{B}_{\vec{k}}=&\left(a^3 \dot{\phi }^2 f' f'' \left(2 \ddot{\phi }+5 H \dot{\phi }\right)+a \left(f'\right)^2 \left(5 a^2 H \dot{\phi } \ddot{\phi }+a^2 \ddot{\phi }^2+\dot{\phi }^2 \left(6 a^2 H^2-4 \dot{H}+k^2\right)\right)+a^3 \dot{\phi }^4 \left(f''\right)^2\right)^{-1}\notag\\&\times f' \left(f' \left(a^2 \left(\left(H \ddot{\phi }+3 \left(4 H^2+\dot{H}\right) \dot{\phi }\right) \tilde{\delta \phi }_{\vec{k}}+4 \dot{\phi } \left(\ddot{\phi }+3 H \dot{\phi }\right) \tilde{\varsigma }_{\vec{k}}\right)+\dot{\phi } \tilde{\varphi }_{\vec{k}} \left(\left(3 a^2-4\right) H \dot{\phi }-a^2 \ddot{\phi }\right)\right)\notag\right.\\&\left.+a^2 \dot{\phi }^2 f'' \left(4 H \tilde{\delta \phi }_{\vec{k}}+\dot{\phi } \left(4 \tilde{\varsigma }_{\vec{k}}-\tilde{\varphi }_{\vec{k}}\right)\right)\right)\,.\label{bk}
\end{align}
Using equation (\ref{bk}), equations (\ref{QS2}) can be written as
\begin{align}
    \tilde{\varsigma}_{\vec{k}}=&-\left(3 a^2 f H \dot{\phi } \ddot{\phi } \left(f'\right)^2-a^2 f \ddot{\phi }^2 \left(f'\right)^2+\dot{\phi }^2 f' \left(-2 a^2 f \ddot{\phi } f''+4 a^2 \ddot{\phi } \left(f'\right)^2+f f' \left(18 a^2 H^2+4 \dot{H}-k^2\right)\right)\notag\right.\\&\left.+3 a^2 H \dot{\phi }^3 f' \left(f f''+4 \left(f'\right)^2\right)+a^2 \dot{\phi }^4 f'' \left(4 \left(f'\right)^2-f f''\right)\right)^{-1}\notag\\&\times \left(a^2 H \dot{\phi } \left(f'\right)^2 \left(\tilde{\delta \phi }_{\vec{k}} \left(\ddot{\phi } f'+24 f H^2+6 f \dot{H}\right)+3 f \ddot{\phi } \tilde{\varphi }_{\vec{k}}\right)+a^2 f \ddot{\phi } \left(f'\right)^2 \left(\ddot{\phi } \tilde{\varphi }_{\vec{k}}+2 H^2 \tilde{\delta \phi }_{\vec{k}}\right)\right.\notag\\&\left.+\dot{\phi }^2 f' \left(\tilde{\varphi }_{\vec{k}} \left(2 a^2 f \ddot{\phi } f''+a^2 \left(-\ddot{\phi }\right) \left(f'\right)^2+f f' \left(4 \left(3 a^2-2\right) H^2-4 \dot{H}+k^2\right)\right)\right.\right.\notag\\&\left.\left.+a^2 \tilde{\delta \phi }_{\vec{k}} \left(8 f H^2 f''+3 \left(4 H^2+\dot{H}\right) \left(f'\right)^2\right)\right)+H \dot{\phi }^3 f' \left(4 a^2 f' f'' \tilde{\delta \phi }_{\vec{k}}+\tilde{\varphi }_{\vec{k}} \left(3 a^2 f f''+\left(3 a^2-4\right) \left(f'\right)^2\right)\right)\right.\notag\\&\left.+a^2 \dot{\phi }^4 f'' \left(f f''-\left(f'\right)^2\right) \tilde{\varphi}_{\vec{k}}\right)\,.\label{vasig}
\end{align}
Using (\ref{bk}) in the equation (\ref{sfe011}), we can express
\begin{align}
    \tilde{\delta \phi}_{\vec{k}}=&-\left(k^2 \left(a^2 f H \dot{\phi} \left(f'\right)^2 \left(3 \left(4 H^2+\dot{H}\right) f'-4 h_0 \ddot{\phi}\right)+a^2 f \ddot{\phi} \left(f'\right)^2 \left(2 h_0 \ddot{\phi}+H^2f'\right)\right.\right.\notag\\&\left.\left.+2 \dot{\phi}^2 f' \left(h_0 \left(2 a^2 f \ddot{\phi} f''-4 a^2 \ddot{\phi} \left(f'\right)^2+f f' \left(-6 a^2 H^2+\left(3 a^2-4\right) \dot{H}+k^2\right)\right)+2 a^2 f H^2 f' f''\right)\right.\right.\notag\\&\left.\left.+2 a^2 h_0 H \dot{\phi}^3 f' \left(f f''-12 \left(f'\right)^2\right)+2 a^2 h_0 \dot{\phi}^4 f'' \left(f f''-4 \left(f'\right)^2\right)\right)\right)^{-1}\notag\\&\times \tilde{\varphi}_{\vec{k}} \left(-4 a^4 f \ddot{\phi}^2 \left(f'\right)^2 \left(h_0 \ddot{\phi}+3 H^2 f'\right)+3 a^2 f H \dot{\phi} \ddot{\phi} \left(f'\right)^3 \left(12 a^2 H^2+k^2\right)\right.\notag\\&\left.+2 \dot{\phi}^4 \left(h_0 \left(-2 a^4 f \ddot{\phi} \left(f''\right)^2+8 a^4 \ddot{\phi} \left(f'\right)^2 f''+72 a^4 H^2 \left(f'\right)^3+3 a^2 f f' f'' \left(6 a^2 H^2+k^2\right)\right)\right.\right.\notag\\&\left.\left.+6 a^4 H^2 f' f'' \left(4 \left(f'\right)^2-f f''\right)\right)\right.\notag\\&+H \dot{\phi}^3 f' \left(2 h_0 \left(-6 a^4 f \ddot{\phi} f''+48 a^4 \ddot{\phi} \left(f'\right)^2+f f' \left(108 a^4 H^2+24 a^2 \dot{H}+9 a^2 k^2-4 k^2\right)\right)\right.\notag\\&\left.\left.+3 a^2 f' \left(f f'' \left(12 a^2 H^2+k^2\right)+48 a^2 H^2 \left(f'\right)^2\right)\right)\right.\notag\\&+\dot{\phi}^2 f' \left(8 a^4 h_0 \ddot{\phi}^2 \left(2 \left(f'\right)^2-f f''\right)+2 a^2 \ddot{\phi} f' \left(12 a^2 H^2 \left(2 \left(f'\right)^2-f f''\right)+f h_0 \left(54 a^2 H^2+8 \dot{H}+k^2\right)\right)\right.\notag\\&\left.\left.+f H^2 \left(f'\right)^2 \left(216 a^4 H^2+48 a^2 \dot{H}+3 a^2 k^2-4 k^2\right)\right)-12 a^4 h_0 H \dot{\phi}^5 f'' \left(f f''-4 \left(f'\right)^2\right)\right)\,,\label{fik}
\end{align}

\begin{align}
G_1=&a^2fH\dot{\phi}\left(f'\right)^2\left(3\left(4H^2+\dot{H}\right)f'-4h_0\ddot{\phi}\right)+a^2f\ddot{\phi}\left(f'\right)^2\left(2h_0\ddot{\phi}+H^2f'\right)\notag\\&+2\dot{\phi}^2f'\left(h_0\left(2a^2f\ddot{\phi}f''-4a^2\ddot{\phi}\left(f'\right)^2+ff'\left(-6a^2H^2+\left(3a^2-4\right)\dot{H}+k^2\right)\right)+2a^2fH^2f'f''\right)\notag\\&+2a^2h_0H\dot{\phi}^3f'\left(ff''-12\left(f'\right)^2\right)+2a^2h_0\dot{\phi}^4f''\left(ff''-4\left(f'\right)^2\right)\,,\\
G_2=&
-2a^4\left(18a^2H^2+k^2\right)h_0^2f''\left(ff''-4\left(f'\right)^2\right)\dot{\phi}^6\notag\\&+4a^4Hh_0\left(27a^2f'f''\left(4\left(f'\right)^2-f f''\right)H^2+h_0\left(6\left(18a^2 H^2+k^2\right)\left(f'\right)^3+24a^2\ddot{\phi}f''\left(f'\right)^2\right.\right.\\&\left.\left.+f\left(27a^2H^2+4k^2\right)f''f'-6a^2f\ddot{\phi}\left(f''\right)^2\right)\right)\dot{\phi}^5
+a^2\left(72a^4\left(f'\right)^2f''\left(4\left(f'\right)^2-ff''\right)H^4\right.\notag\\&\left.+h_0\left(1296H^4\left(f'\right)^4a^4-36fH^2\ddot{\phi}f'\left(f''\right)^2a^4+144H^2\ddot{\phi}\left(f'\right)^3f''a^4+4f^2k^2\left(3a^2H^2+k^2\right)\left(f''\right)^2\right.\right.\notag\\&\left.\left.+f\left(324a^4H^4+65a^2k^2H^2-7k^4\right)\left(f'\right)^2f''\right)+2h_0^2\left(-2f\ddot{\phi}^2\left(f''\right)^2a^4+4\left(54a^2H^2+k^2\right) \ddot{\phi}\left(f'\right)^3a^2\right.\right.\notag\\&\left.\left.+fk^2\ddot{\phi}f'f''a^2+\left(f'\right)^2\left(8\ddot{\phi}^2f''a^4+f\left(324a^4H^4+33a^2k^2H^2-12k^2H^2-k^4\right)+f\left(\left(72H^2-3k^2\right)a^2+4k^2\right)\dot{H}\right)\right)\right)\dot{\phi}^4\notag\\&
+Hf'\left(36\ddot{\phi}^2h_0^2\left(4\left(f'\right)^2-ff''\right)a^6+18H^2\left(f'\right)^2\left(48a^2H^2\left(f'\right)^2+f\left(12a^2H^2+5k^2\right)f''\right)a^4\right.\notag\\&\left.+\ddot{\phi}h_0f'\left(864H^2\left(f'\right)^2a^4+27f\left(k^2-4a^2H^2\right)f''a^2+4f\left(135H^2a^4+7k^2a^2+24\dot{H}a^2-2k^2\right)h_0\right)a^2\right.\notag\\&\left.+fh_0\left(4a^2f\left(27a^2H^2+4k^2\right)f''k^2\right.\right.\notag\\&\left.\left.+3\left(648 H^4a^6+59H^2k^2a^4+\left(144H^2+17k^2\right)\dot{H}a^4-\left(k^4+20H^2k^2\right)a^2-4k^4\right)\left(f'\right)^2\right)\right)\dot{\phi}^3\notag\\&
+f'\left(4\ddot{\phi}^2h_0^2\left(4\ddot{\phi}\left(f'\right)^2a^2-2f\ddot{\phi}f''a^2+f\left(27a^2H^2+4\dot{H}\right)f'\right)a^4\right.\notag\\&\left.+2H^2f'\left(144H^2\ddot{\phi} \left(f'\right)^3a^4-72fH^2\ddot{\phi}f'f''a^4+3f\left(3\left(16H^2+3k^2\right)\dot{H}a^2+H^2\left(216H^2a^4+39k^2a^2-4k^2\right)\right) \left(f'\right)^2\right.\right.\notag\\&\left.+4f^2k^2\left(15a^2H^2+k^2\right)f''\right)a^2\notag\\&+h_0\left(144H^2\ddot{\phi}^2\left(f'\right)^3a^6+f\left(972a^4H^4+65a^2 k^2H^2-4k^2H^2-7k^4+18a^2\left(8H^2+k^2\right)\dot{H}\right)\ddot{\phi}\left(f'\right)^2a^2\right.\notag\\&\left.+8f^2k^2\left(7a^2H^2+k^2\right)\ddot{\phi}f''a^2+4ff'\left(-18H^2\ddot{\phi}^2f''a^6+fk^2\left(54a^4H^4+12a^2k^2H^2-4k^2H^2+k^4\right)\right.\right.\notag\\&\left.\left.\left.+fk^2\left(27H^2a^4+3\left(k^2-4H^2\right)a^2-4k^2\right)\dot{H}\right)\right)\right)\dot{\phi}^2\notag\\&
+a^2fH\left(f'\right)^2\left(-12\ddot{\phi}^3h_0^2a^4+9k^2\ddot{\phi}^2h_0f'a^2+6fk^2\left(15a^2H^2+k^2\right)\left(4H^2+\dot{H}\right)f'\right.\notag\\&\left.+4\ddot{\phi}\left(f\left(24H^2a^2+6\dot{H}a^2+k^2\right)h_0k^2+9a^2H^2\left(6a^2H^2+k^2\right)\left(f'\right)^2\right)\right)\dot{\phi}\notag\\&+2a^2f\ddot{\phi}\left(f'\right)^2\left(-2\ddot{\phi}^3h_0^2a^4-18H^2\ddot{\phi}^2h_0f'a^4+fH^2k^2\left(15a^2H^2+k^2\right)f'+\ddot{\phi}\left(2fk^2\left(5a^2H^2+k^2\right)h_0-36a^4H^4\left(f'\right)^2\right)\right)\,.
\end{align}
\end{document}